\documentclass[nobibnotes,final,5p,times,twocolumn,authoryear,numbers, square, sort&compress]{elsarticle}

\usepackage{epsfig}

\usepackage{amssymb}
\usepackage{lipsum}
\usepackage{amsmath}
\usepackage{amsthm}

\usepackage[hidelinks]{hyperref}

\usepackage{graphicx,framed} 
\usepackage{dcolumn} 
\usepackage{bm} 
\usepackage{braket}
\usepackage{dsfont}
\usepackage{color}
\usepackage{amsthm}
\usepackage[normalem]{ulem} 
\usepackage{qcircuit}       

\usepackage{amsthm}

\usepackage{comment}
\usepackage{float}
\setcitestyle{numbers}
\usepackage{flushend}

\makeatletter
\def\ps@pprintTitle{%
     \let\@oddhead\@empty
     \let\@evenhead\@empty
     \def\@oddfoot{\reset@font\hfil}%
     \let\@evenfoot\@oddfoot
}
\makeatother

\begin{document}

\begin{frontmatter}



\title{Testing Bell-CHSH Inequalities Using topological\\ Aharonov-Casher and He-McKellar-Wilkens Phases}


\author{H. O. Cildiroglu}
\affiliation{organization={Boston University },
            addressline={Physics Department}, 
            city={Boston},
            postcode={02100}, 
            state={MA},
            country={USA}}
\affiliation{organization={Ankara University },
            addressline={Department of Physics Engineering}, 
            postcode={06100}, 
            state={Ankara},
            country={Türkiye}}

\begin{abstract}
The effects of Aharonov-Casher (AC) and He-McKellar-Wilkens (HMW) phases on entangled spin-1/2 quantum systems are investigated. We use linear charge distributions positioned at the center of resulting closed orbits, capitalizing on Mach Zender-type interferometers modified with phase retarders to unveil the topological effects. We analyze how AC-HMW phases influence the Bell angles and maximal violation of Bell-CHSH inequalities (BI) without any classical interaction. We incorporate the spin and path of particles in the interferometers as observables to test noncontextual hidden variable theories against quantum mechanics, leveraging the non-local features of AC-HMW effects. Furthermore, we discuss potential implementations of our scheme in physical systems.

\end{abstract}




\end{frontmatter}




\section{Introduction}
\label{introduction}

Testing Bell-CHSH inequalities (BI) which impose constraints on the correlations observed among distinct components within multipartite systems within the framework of local realism is crucial for serving as experimental validations of the foundational principles of quantum mechanics, particularly the concept of quantum entanglement ~\cite{Einstein1935, Bell1964, Clauser1969}. BI has been tested in diverse quantum systems such as photons and atoms in quantum optics, high-energy physics, neutrals, charmonium decays, neutrino oscillations ~\cite{Aspect1982, Ou1988, Horodecki1995, Weihs1998, Brendel1999, Benatti2000, Acin2001, Genovese2005, Ursin2007, Banerjee2015, Qian2020, atlas2023observation, guedes2024unruh}, and the experiments indicate the existence of non-local features of quantum mechanics.

On the other hand, the topological properties of physical systems are crucial for quantum communication and quantum computation applications with their non-local features. One of the best-known and earliest examples of this is the Aharonov-Bohm (AB) physical/quantum mechanical process ~\cite{Aharonov1959, Chambers1960, Tonomura1986}. AB-type effects are explained by the introduction of the vector potential or vector potential-like physical quantities as complex phase factors into the wave functions of particles moving along the closed trajectories around singular regions created by electromagnetic field sources without any classical interactions. Also, AB-type effects are invariant under spatial translations through the z direction (traditionally) due to the symmetries of the problems, and there are duality and identity relationships between phases particularly appearing in two dimensions ~\cite{Dowling1999, Cildiroglu2021}. Accordingly, while AB and its fully dual-AB (DAB) effects are independent of the spin orientations of moving particles, Aharonov-Casher (AC) and its fully dual He-McKellar-Wilkens (HMW) effects demonstrate dependence on spin polarizations ~\cite{Aharonov1984, Cimmino1989, He1993, Wilkens1994, Gillot2013, Cildiroglu2015}. Between AB and AC (similarly between HMW and DAB) one can speak of an identity rather than a duality that arises in the static reference frames of charges, especially if the dipoles are polarized. In this regard, AC-HMW effects serve as practical tools for testing nonlocality and noncontextuality in comparison to quantum mechanics within EPR-Bohm-type experiments ~\cite{Pati1998, Bertlmann2004, Su2013, Cirelson1980}.

\begin{figure*}
\centering 
\includegraphics[width=0.7\textwidth]{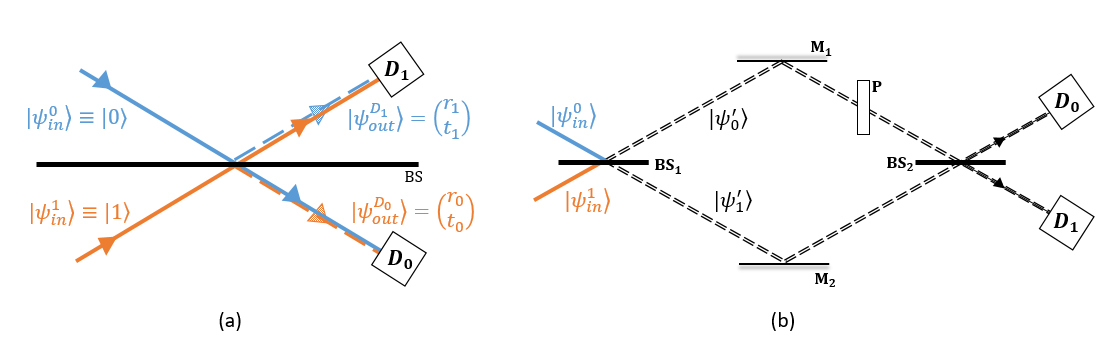}
\caption{\label{fig:wide2} (a) Two ports symmetric and lossless (50:50) BS. (b) Schematic representation of the BS-P-BS system}
\end{figure*}

In this work, we propose a scheme to observe the effects of topological phases on entangled quantum states by exploiting the non-local structure of the AC phase and to test of incompatibility of noncontextuality with quantum mechanics, which is a restrictive demand for a theory, via BI without any classical interaction or adiabaticity condition. For this purpose, we use a pair of Mach Zender-type interferometers modified with phase retarders to be analogous to spin measurement angles in arbitrary directions (BS-P-BS). We place the electromagnetic field source (linear charge distributions) in the closed regions between the legs of the interferometers, which creates the topological phase contributions to the wave functions of the moving particles (dipoles). We demonstrate that the correlation function becomes dependent on the topological phase by obtaining the detection probabilities (or spin measurement probabilities) for fixed phases of retarders (or Bell angles) maximally violated by quantum mechanical expectation values after passing through mutual BS-P-BS systems. 

This letter is organized as follows. In the second section, we provide the background of the study as a preliminary preparation. We review the quantum mechanical use of Beam-Splitters (BS) with phase retarders (P) and consider the possible scenarios for BI testing for all kinds of quantum mechanical particles (quantons), not only photons, by the existing notation. In the third section, we first modify the proposed scenarios with the contribution of the AC phase. Emphasizing the importance of spin dependences of moving particles, we obtain the correlation function for testing BI in the BS-P-BS system that becomes AC-HMW phase dependent. Lastly, we discuss the possible implementations of our scheme in physical systems and a generalization that can be used for other studies in the literature.

\section{Notation and Background}

Beam splitters (BS) are the most crucial components of theoretical or experimental studies that demonstrate the behaviors of photons as particles and the existence of physical properties like non-locality or indistinguishability. The main purpose of using BS appears to be  straightforward from the classical perspective; however, its significance in quantum physics is far more profound ~\cite{Zeilinger1981}. Moreover, special properties emerge when BSs are utilized in combination with phase retarders (P) in Mach-Zender interferometers (BS-P-BS) revealing strong connections between spin measurement and detection of the quantum mechanical particles (quantons). Accordingly, the probabilities of being spin up or down in an arbitrary direction are equivalent to the probabilities of quantons being detected by detectors. The phase of the retarder corresponds to the angle of the direction in which the spin of quantons is measured. In this way, they can be easily used in systems consisting of two spatially correlated (or similarly spin-entangled) quantons in EPR-Bohm-type experiments ~\cite{Bohm1957}. Besides, the BS-P-BS structure that contains closed trajectories for each quanton allows us to study other quantum mechanical/physical effects that seem to have common origins.  Hence, the effects of topological phases on spatially correlated systems can be investigated by introducing electromagnetic field sources at the singular region where the quantons are not allowed to move. In this way, the detection probabilities can be achieved for spin-entangled quantons to find the expectation values and correlation functions. Thus, it can be used for testing BI without photons ~\cite{Bell1964, Clauser1969}.

Lets consider a symmetric and lossless  BS with two ports $0$ and $1$ for incoming, and two ports $D_0$ and $D_1$ for outgoing quantons (See Fig.~\ref{fig:wide2}.a). The basis $(1\thinspace,0)^T\equiv\left|0\right\rangle$ and $(0\thinspace,1)^T\equiv\left|1\right\rangle$ for ports 0 and 1 can be used to describe the state $\left|\psi^i_{in}\right\rangle$ of incoming quantons entering such a BS through the port $i$ $(i=0,1)$. Similarly, to represent the output states one can choose the basis $(1\thinspace,0)^T$ and $(0\thinspace,1)^T$ for $\left|D_{0}\right\rangle$ and $\left|D_{1}\right\rangle$. Hence, the BS performs a linear transformation that converts $\left|\psi^i_{in}\right\rangle$ to $\left|\psi_{out}\right\rangle$. By denoting purely real $(r_{i}=R_{i})$ and purely imaginary $(t_{i}=iT_{i})$ as reflection and transmission amplitudes of the BS respectively (Namely, for incoming quantons through port 0 as $r_{0}$ and $t_{0}$, and  through port 1 as $r_{1}$ and $t_{1}$), $\left|r_{0}\right|=\left|r_{1}\right|=\left|t_{0}\right|=\left|t_{1}\right|$; $\left|r_{i}\right|^{2}+\left|t_{i}\right|^{2}=1$; $r_{0}^{*}t_{1}+t_{0}^{*}r_{1}=0$ are provided. Therefore the operation of the BS can be given by a unitary matrix, 

\begin{equation}
\label{eq:1}
BS=\left(\begin{array}{cc} r_{0} & t_{1}\\
t_{0} & r_{1} \end{array}\right)=\frac{1}{\sqrt{2}}\left(\begin{array}{cc} 1 & i\\
i & 1 \end{array}\right)
\end{equation}

\begin{figure*}
\centering 
\includegraphics[width=0.7\linewidth]{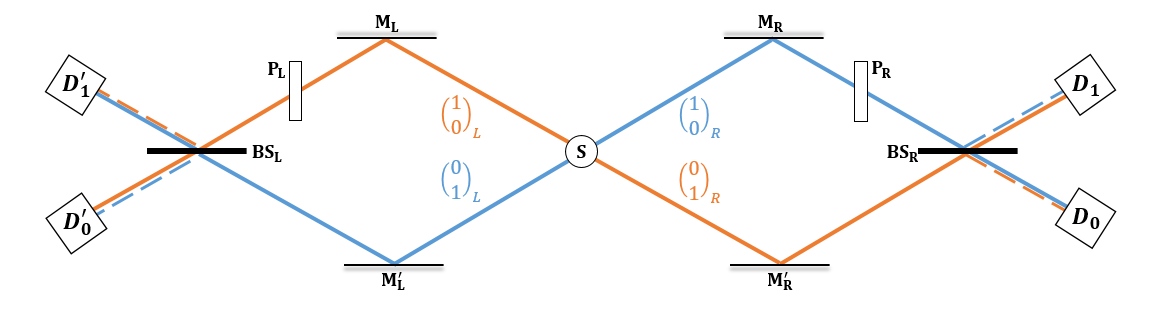}
\caption{\label{fig:wide3} A scheme for the use of BS-P for spatially correlated two-quanton systems generated from the same source}
\end{figure*}
 
\noindent Thus, in the case of single incoming quanton through port $i$, the output state is $\left|\psi^{0(1)}_{in}\right\rangle \rightarrow\left|\psi_{out}\right\rangle=\frac{1}{\sqrt{2}}\left[\left|D_{0(1)}\right\rangle +i\left|D_{1(0)}\right\rangle\right]$, and the detection probabilities are 1/2 for each detector. 

On the other hand, some significant characteristics appear when a second BS is added with a phase retarder. The detection probabilities of quantons become dependent on the phase of the retarder (See Fig.~\ref{fig:wide2}.b). Furthermore, this phase corresponds to the angle of the arbitrary direction on which spin measurements are performed to find the probabilities (up or down). According to the existing configuration, wlog, the retarder operator ${P}$ can be expressed as,

\begin{equation}
\label{eq:2}
P=\left(\begin{array}{cc} 
e^{i\vartheta} & 0\\
0 & 1
\end{array}\right)\thinspace\thinspace    
\end{equation}

\noindent In the case of single incoming quanton through port $i$, system evolves into the state $\left|\psi^i_{in}\right\rangle \rightarrow \left|\psi_{out}\right\rangle = {[BS_{2}][P][BS_{1}]}\left|\psi^i_{in}\right\rangle=\frac{e^{i\vartheta}}{\sqrt{2}}\left[i\left|{D_{0}}\right\rangle +\left|{D_{1}}\right\rangle \right]$. Consequently, the final state of the system can be given in a compact representation,

\begin{equation}
\label{eq:3}
\left(\begin{array}{cc}
\left|0\right\rangle\\
\left|1\right\rangle
\end{array}\right)
\rightarrow\left|\psi_{out}\right\rangle= \frac{ie^{i\frac{\vartheta}{2}}}{\sqrt{2}}
\left(\begin{array}{cc}
-\sin{\frac{\vartheta}{2}} & \cos{\frac{\vartheta}{2}}\\
\cos{\frac{\vartheta}{2}} & \sin{\frac{\vartheta}{2}}
\end{array}\right)
\left(\begin{array}{cc}
\left|{D_{0}}\right\rangle\\
\left|{D_{1}}\right\rangle
\end{array}\right)
\end{equation}

\noindent The term $(ie^{i\frac{\vartheta}{2}})$ is the common phase factor and can be neglected for both two equations. These results are valid for all kinds of quantons (bosons, fermions) and just related to the spatial part of the particle’s wavefunctions. 

Now, we extend the problem in correlated systems in two scenarios.  First, let's assume that two spatially correlated quantons are produced from the same source, and follow either orange or blue lines as in Fig.~\ref{fig:wide3}. In this scenario, the initial state of the system can be considered as,

\begin{equation}
\label{eq:4}
\left|\psi_{in}\right\rangle=\frac{1}{\sqrt{2}}\left[\left|0\right\rangle_L\otimes\left|1\right\rangle_R-\left|1\right\rangle_L\otimes\left|0\right\rangle_R\right]
\end{equation}

\noindent and one can use ~\eqref{eq:2} as retarder operator with the selection of the phase $\vartheta_{L(R)}$ consistent with the path. The correlated quantons reach BSs after passing through reflecting mirrors and retarders on both sides. Thus, the state of the system ~\eqref{eq:4} transformed into the state $\left|\psi_{out}\right\rangle = (BS_{L}\otimes BS_{R})(P_{L}\otimes P_{R})\left|\psi_{in}\right\rangle$. It can be written in an explicit form by neglecting the overall phase factor,

\begin{equation}
\begin{aligned}
\label{eq:5}
\left|\psi_{out}\right\rangle &= \frac{1}{2\sqrt{2}} ([\thinspace|{D_{0}}'\rangle+i|{D_{1}}'\rangle] \otimes [i|{D_{0}} \rangle+ |{D_{1}}\rangle] \\
&\quad - e^{i\vartheta} [\thinspace i|{D_{0}}'\rangle+|{D_{1}}'\rangle] \otimes [\thinspace |{D_{0}}\rangle+i|{D_{1}}\rangle])
\end{aligned}
\end{equation}

\noindent where $\vartheta_{R}-\vartheta_{L}=\vartheta$. At this stage, joint-detection amplitudes for a quanton at e.g. $D_0$ and the other one at $D'_{0}$ (or $D'_{1}$),

\begin{equation}
\begin{aligned}
\label{eq:6}
A(D_0^{'},D_{0}) & \equiv \left\langle {D_0^{'}} {D_{0}}\right|\left.\psi_{out}\right\rangle = \frac{1}{2\sqrt{2}}i(1-e^{i\vartheta}) \\
A(D'_{1}, D_{0}) & \equiv \left\langle {D'_{1}}{D_{0}}\right|\left.\psi_{out}\right\rangle = -\frac{1}{2\sqrt{2}}(1+e^{i\vartheta})
\end{aligned}
\end{equation}

\noindent and the associated probabilities are revealed as:

\begin{equation}
\begin{aligned}
\label{eq:7}
P(D_0^{'},D_{0}) & \equiv \left\Vert A(D_0^{'},D_{0})\right\Vert ^{2}= \frac{1}{2}\cos^2{\left(\frac{\vartheta_L-\vartheta_R}{2}\right)} \\
P(D_1^{'},D_{0}) & \equiv \left\Vert A(D_1^{'},D_{0})\right\Vert ^{2}= \frac{1}{2} \sin^2\left(\frac{\vartheta_L-\vartheta_R}{2}\right).
\end{aligned}
\end{equation}

These results are related to the joint-spin measurement probabilities obtained in EPR-Bohm type experimental setups using spin-correlated particles, with a phase shift $\pi/2$ as Degiorgio's relation due to BS ~\cite{Degiorgio1980}. It is also clear that the probabilities depend on the phase of the retarder. 

\begin{figure*}
\centering 
\includegraphics[width=0.7\textwidth]{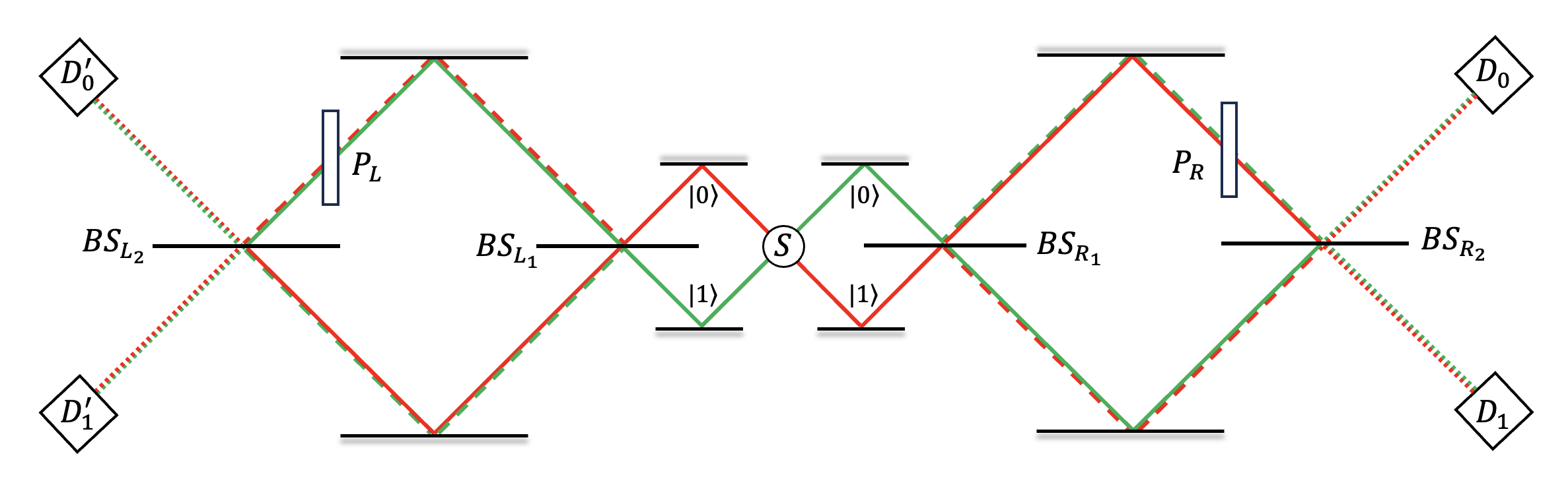}
\caption{\label{fig:wide4} The use of the BS-P-BS system based on detection probabilities as an analog to spin measurement probabilities for spatially correlated two-quanton systems and a scheme for a gedanken experimental setup for testing BI.}
\end{figure*}

Second, BS-P-BS system can be proposed to establish an analogy between spin measurements and detection processes. Accordingly, spatially correlated quantons produced in the state ~\eqref{eq:4} are subjected to retarders placed on one of the arms after passing the first BSs. Then, they move through the second BSs and reach the detectors (See Fig.~\ref{fig:wide4}). In this new configuration, the final state of the system is obtained as $\left|\psi_{out}\right\rangle = (BS_{L_1}\otimes BS_{R_1})(P_{L}\otimes P_{R})(BS_{L_2}\otimes BS_{R_2})\left|\psi_{in}\right\rangle$, or clearly,

\begin{align}
\label{eq:8}
& |\psi_{out}\rangle = \frac{1}{\sqrt{2}}[(-\sin{\frac{\vartheta_L}{2}}|{D_{0}}'\rangle+\cos{\frac{\vartheta_L}{2}}|{D_{1}}'\rangle) \nonumber \otimes \\ &(\cos{\frac{\vartheta_R}{2}} |D_{0}\rangle + \sin{\frac{\vartheta_R}{2}}|D_{1}\rangle) -(\cos{\frac{\vartheta_L}{2}}|D_{0}'\rangle + \sin{\frac{\vartheta_L}{2}}|D_{1}'\rangle) \nonumber \\ 
& \thinspace\thinspace\thinspace\thinspace \thinspace\thinspace\thinspace\thinspace\thinspace\thinspace\thinspace \thinspace\thinspace\thinspace\thinspace\thinspace\thinspace \otimes (-\sin{\frac{\vartheta_R}{2}}|D_{0}\rangle + \cos{\frac{\vartheta_R}{2}}|D_{1}\rangle)]
\end{align}

\noindent Thus, one can find the joint-detection probabilities for quantons at $D_{0}$ and $D_0'$ (or $D_{1}'$):

\begin{eqnarray}
\label{eq:9}
P(D_0^{'},D_{0}) = \frac{1}{2} \sin^2\left(\frac{\vartheta_L - \vartheta_R}{2}\right) \equiv P(\vartheta_L\Uparrow_{\tilde{n}},\vartheta_R \Uparrow_{\tilde{n}})  \nonumber \\
P(D_1^{'},D_{0}) = \frac{1}{2}\cos^2{\left(\frac{\vartheta_L - \vartheta_R}{2}\right)} \equiv P(\vartheta_L \Downarrow_{\tilde{n}},\vartheta_R \Uparrow_{\tilde{n}})
\end{eqnarray}

The above results are well-known expressions in entangled quantum systems ~\cite{Cildiroglu2021}. The retarder phase corresponds to the spin measurement angle $\hat{\boldsymbol{n}}$. Equation ~\eqref{eq:9} clearly connects these processes without any Degiorgio phase shift. Accordingly, the probabilities of spin up (or down) in $\hat{\boldsymbol{n}}$ direction are equivalent to the probabilities detected by a detector placed up (or down) for the corresponding phase of the retarder to the angle of n with horizontal. Moreover, this setup can be used for testing BI, crucial for demonstrating non-local properties of quantum mechanics. Therefore, in EPR-Bohm-type experimental setups, one tries to find the relevant correlation function S as a combination of expectation value expressions based on different spin-measurement probabilities. In general, to obtain the correlation function S in entangled two spin-1/2 particles in a singlet state,

\begin{equation}
\label{eq:10}
\left|\psi_{in}\right\rangle=\frac{1}{\sqrt{2}}\left(\left|\Uparrow_{\boldsymbol{\hat{n}}}\right\rangle_L\otimes\left|\Downarrow_{\boldsymbol{\hat{n}}}\right\rangle_R-\left|\Downarrow_{\boldsymbol{\hat{n}}}\right\rangle_L\otimes\left|\Uparrow_{\boldsymbol{\hat{n}}}\right\rangle_R\right)
\end{equation}

\noindent One can start with decomposing it into the instantaneous eigenstates of the system in any arbitrarily chosen direction $\boldsymbol{\hat{n}}$, then observe the joint-spin measurement probabilities of particles in the directions that make different $\vartheta_L$ and $\vartheta_R$ angles with $\boldsymbol{\hat{n}}$ on both sides. Here, the spin measurement probabilities are replaced by the detection probabilities. In other words, $P\left(\boldsymbol{\vartheta_L}\Uparrow_{\boldsymbol{\hat{n}}},\boldsymbol{\vartheta_R}\Uparrow_{\boldsymbol{\hat{n}}}\right)$ and $P\left(\boldsymbol{\vartheta_L} \Uparrow_{\boldsymbol{\hat{n}}},\boldsymbol{\vartheta_R}\Downarrow_{\boldsymbol{\hat{n}}} \right)$ are exactly same with $P(D_0^{'},D_{0})$ and $P(D'_{1}, D_{0})$. By introducing the observable $\mathcal{O}$,

\begin{equation}
\label{eq:11}
\mathcal{O}^{l(r)}(\vartheta_L(R))=P_{D_0^{'}}^{l(r)}(\vartheta_L(R) )-P_{D_1^{'}}^{l(r)}(\vartheta_L(R))
\end{equation}

\noindent one can calculate the expectation values, 

\begin{equation}
\label{eq:12}
    E(\vartheta_L, \vartheta_R) = \left\langle \psi\left(\tau\right)\right|\mathcal{O}^{l}(\vartheta_L)\otimes \mathcal{O}^{r}(\vartheta_R)\left|\psi\left(\tau\right)\right\rangle = -\cos{(\vartheta_L-\vartheta_R)}
\end{equation}

\noindent and the correlation function can be derived to test the BI as,

\begin{eqnarray}
\label{eq:13}
S(\vartheta_L, \vartheta_R, \vartheta_L',\vartheta_R')=\left|E(\vartheta_L,\vartheta_R)-E(\vartheta_L,\vartheta_L') \right|\nonumber\\
+\left|E(\vartheta_R',\vartheta_R)+E(\vartheta_R',\vartheta_L') \right|
\end{eqnarray}

\noindent With the certain chosen angles for the correlation function $ S(0,\frac{\pi }{4},\frac{3\pi }{4},\frac{\pi}{2})=2\sqrt{2}>2$, BI is maximally violated by the quantum mechanical expectation values.  
\section{Testing Bell-CHSH inequalities using topological phases}


Topological phases are quantum mechanical phenomena that occur in multiply-connected regions under the influence of vector potential or vector potential-like physical quantities, without classical Lorentz forces, independent of the geometry of the space and the speed of the particles ~\cite{Aharonov1959, Chambers1960, Tonomura1986}. It is also invariant under spatial translations (traditionally in the z-direction) and essentially arises in two-dimensional space. The first and most well-known example of topological phases is the AB phase, which is based on the control of the interference pattern via an infinitely long and thin solenoid placed beyond the reach of moving electrons in a double-slit experiment. Subsequently, consequential studies have been performed and experimentally observed, including duality and identity relations with the AB system ~\cite{Dowling1999, Cildiroglu2021, Aharonov1984, Cimmino1989, He1993, Wilkens1994, Gillot2013, Cildiroglu2015, Silverman1990, Delgado2008, Maciejko2010, Singleton2013, Bright2015, Singleton2016, Jing2017, Cildiroglu2018, 1Hashemi2018, Choudhury2019, Jing2020, Tunalioglu2023, Saldanha2023, Wakamatsu2024}. At the forefront of these is the AC effect arises from the motion of chargeless neutrons carrying a magnetic dipole in closed trajectories around an electrically linear charge distribution ~\cite{Aharonov1984, Cimmino1989}. Similarly, the HMW effect occurs due to the motion of chargeless particles carrying an electric dipole around a linear charge distribution acting as a magnetic field source ~\cite{He1993, Wilkens1994, Gillot2013}. The AC and HMW Hamiltonians describing the interaction and the transformed wave functions are,

\begin{equation}
\label{eq:14}
\begin{matrix}
    \Delta H_{AC}=-s \boldsymbol{\mu} \cdot\boldsymbol{\tilde{E}} & & & \Delta H_{HMW}=-s \boldsymbol{d} \cdot\boldsymbol{\tilde{B}} \\
    \psi' =e^{-is\mu \lambda_E }\psi_0 & & & \psi' =e^{-isd \lambda_B }\psi_0
\end{matrix}
\end{equation}

\begin{figure*}
\includegraphics[width=1.0\textwidth]{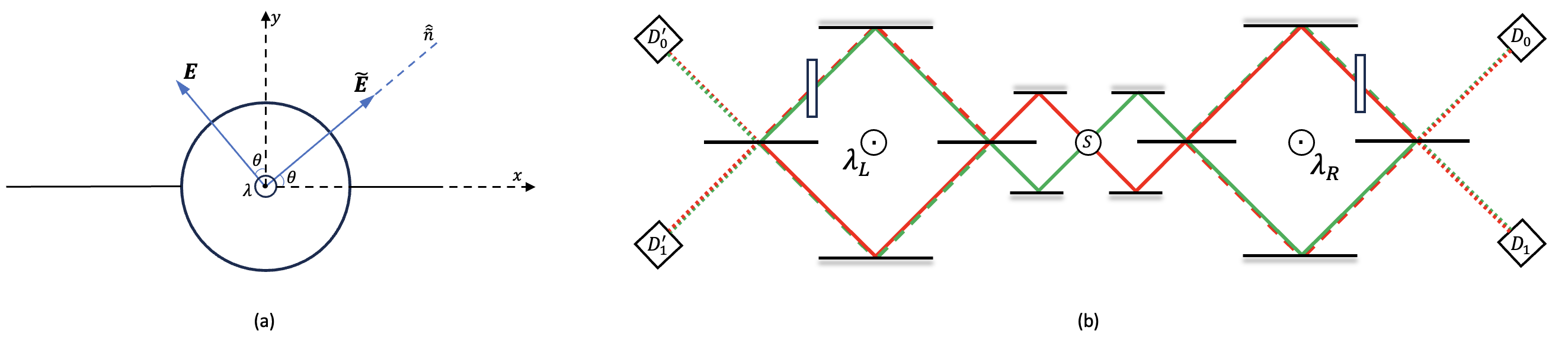}
\caption{\label{fig:wide5} (a) Schematic representation of the AC phase in 2+1 dimensions resulting from the motion of magnetic dipole moment carrier chargeless particles (or HMW phase in the case of $\boldsymbol{\tilde{E}} \longrightarrow \boldsymbol{\tilde{B}}$ and $\mu \longrightarrow d$) in a closed trajectory around a linear charge distribution. (b) A proposal for a gedanken experimental setup to investigate the effects of topological phases on entangled quantum states and to observe the violations of BI using physical quantities in the singular region to test the non-local properties of quantum mechanics.}
\end{figure*}

\noindent Here, $\boldsymbol{\mu}$ and $\boldsymbol{d}$ are the magnetic and  electric dipoles respectively, $\boldsymbol{\tilde{E}}=\boldsymbol{E}\times \boldsymbol{\hat{z}}$ and $\boldsymbol{\tilde{B}}=\boldsymbol{B}\times \boldsymbol{\hat{z}}$ denote the confined electric and magnetic fields in two dimensions, $\lambda_E$ and $\lambda_B$ correspond to the electric and magnetic linear charge distributions respectively ($\lambda_{E(B)}=\oint \boldsymbol{\Tilde{E}}(\boldsymbol{\Tilde{B}}) \cdot d\boldsymbol{l}$), the parameter $s=\pm 1$ arises from the two inequivalent representations of Dirac algebra in two dimensions, corresponding to spin up and down particles (for further details see ~\cite{Cildiroglu2021}). The expression ~\eqref{eq:14} clearly demonstrates the AC and HMW dualities. Henceforth, we use a representation based on AC, suggesting simultaneous transformations $\mu \longrightarrow d$ and $\boldsymbol{\tilde{E}} \longrightarrow \boldsymbol{\tilde{B}}$ for the HMW effect for the sake of simplicity. The instantaneous eigenstates of the system under interactions along the $\boldsymbol{\hat{\tilde{n}}}$ direction (or $\boldsymbol{\tilde{E}}$ direction) are,

\begin{equation}
\label{eq:15}
\begin{matrix}\left|\Uparrow_{\tilde{n}};t\right\rangle =\frac{1}{\sqrt{2}}(-e^{-i\theta },1)^{T}\\
\\
\left|\Downarrow_{\tilde{n}};t\right\rangle =\frac{1}{\sqrt{2}}(e^{-i\theta},1)^{T}
\end{matrix}\end{equation}

\noindent and the corresponding eigenvalues are $\lambda _{AC_{1,2}}= s \mu \tilde{E}$ with the definitions $\tilde{E}_{x}=\tilde{E}\cos{\theta} , \tilde{E}_{y}=\tilde{E} \sin{\theta}$ and $\tilde{E}^{2}=\tilde{E}_{x}^{2}+\tilde{E}_{y}^{2}$,  where $\theta $ is the angle between positive $x$ axis and arbitrary $\boldsymbol{\hat{\tilde{n}}}$ direction (See Fig.~\ref{fig:wide5}.a).  Each eigenstate gains a topological AC phase associated with their spin orientation after a complete cycle at $t=\tau $, 

\begin{equation}
\label{eq:16}
\begin{matrix}
\left|\Uparrow_{\tilde{n}};t=0\right\rangle \to  \left|\Uparrow_{\tilde{n}};t=\tau\right\rangle = e^{-i\mu\lambda}\left|\Uparrow_{\tilde{n}};t=0\right\rangle & \\& \\
\left|\Downarrow_{\tilde{n}};t=0\right\rangle \to \left|\Downarrow_{\tilde{n}};t=\tau\right\rangle = e^{i\mu\lambda}\left|\Downarrow_{\tilde{n}};t=0\right\rangle
\end{matrix}
\end{equation}

\noindent At this stage, above equations provide a suitable basis for the study of entangled quantum states, and allows the setups proposed in the previous section to be developed with the AC phase. 

Now, first let us verify that the effect of the AC phase cannot be observed with a source of the electrical field linear charge distribution placed in the setup given in Fig.~\ref{fig:wide3}. With the selection of initial state ~\eqref{eq:4} for the spatially-correlated (or spin-entangled) particle pair, the final state of the system is 
$\left|\psi_{out}\right\rangle' = ({BS_{L}} \otimes {BS_{R}})({T_{L}}\otimes {T_{R}})(P_{L}\otimes P_{R}) \left|\psi_{in}\right\rangle$. Here, $T_{L(R)}$ is the quantum topological phase operator,

\begin{equation}
\label{eq:17}
T_{L(R)}=\left(\begin{array}{cc} 
e^{i\mu \int_{u} \boldsymbol{\Tilde{E}}_{L(R)}\cdot d\boldsymbol{l}} & 0\\
0 & e^{-i\mu \int_{d} \boldsymbol{\Tilde{E}}_{L(R)}\cdot d\boldsymbol{l}} 
\end{array}\right)\thinspace\thinspace
\end{equation}

\noindent Thus, the final state of the system is obtained with a complex phase contribution to the output state in ~\eqref{eq:4} by neglecting the rest overall phase factors with the definition for particles following upper (u) and lower (d) trajectories $(\int_{u} \boldsymbol{\Tilde{E}}_{L(R)}\cdot d\boldsymbol{l} - \int_{d} \boldsymbol{\Tilde{E}}_{L(R)}\cdot d\boldsymbol{l}=\int_{u} \boldsymbol{\Tilde{E}}_{L(R)}\cdot d\boldsymbol{l} + \int_{d'} \boldsymbol{\Tilde{E}}_{L(R)}\cdot d\boldsymbol{l}=\oint \boldsymbol{\Tilde{E}}_{L(R)}\cdot d\boldsymbol{l}=\lambda_{L(R)})$.

\begin{equation}
\label{eq:18}
\left|\psi_{out}\right\rangle'=e^{i\mu\Delta\lambda} \left|\psi^{(4)}_{out}\right\rangle
\end{equation}

\noindent The probabilities of detection (or spin measurements) for quantons are the same as ~\eqref{eq:7}. Hence, the overall phase factor disappears in the expectation value expressions as predicted. 

Second, to observe such effects, the physical/quantum mechanical processes in Fig.~\ref{fig:wide4} can be improved by adding linear charge distributions (See Fig.~\ref{fig:wide5}.b). Similarly, by choosing ~\eqref{eq:4} as the initial state, the output state of the system is $\left|\psi_{out}\right\rangle = ({BS_{L}}_2 \otimes {BS_{R}}_2)({T_{L}} \otimes {T_{R}})(P_{L}\otimes P_{R})({BS_{L}}_1 \otimes {BS_{R}}_1) \left|\psi_{in}\right\rangle$. By neglecting the overall phase factor, and taking into consideration ~\eqref{eq:2}, ~\eqref{eq:3} and ~\eqref{eq:17}, the final state of the system can be rewritten explicitly as ($\lambda = \lambda_L - \lambda_R$),

\begin{align}
\label{eq:19}
|\psi_{out}\rangle &= \frac{1}{\sqrt{2}}[(-\sin{\frac{\vartheta_L}{2}}|{D_{0}}'\rangle+\cos{\frac{\vartheta_L}{2}}|{D_{1}}'\rangle) \otimes  (\cos{\frac{\vartheta_R}{2}} |D_{0}\rangle \nonumber \\ & + \sin{\frac{\vartheta_R}{2}}|D_{1}\rangle) - e^{-2i\mu\lambda}(\cos{\frac{\vartheta_L}{2}}|D_{0}'\rangle + \sin{\frac{\vartheta_L}{2}}|D_{1}'\rangle) \\ & \otimes (-\sin{\frac{\vartheta_R}{2}}|D_{0}\rangle + \cos{\frac{\vartheta_R}{2}}|D_{1}\rangle)] \nonumber
\end{align}

\noindent The phase $(2\mu\lambda)$ in ~\eqref{eq:19} appears in the expectation value expressions in contrast to ~\eqref{eq:18}, and it allows the correlation function S to be controlled by a physical quantity placed in the singular region without any classical interaction. To demonstrate this, we first find the joint-detection probabilities (or analogously the joint-spin measurement probabilities for arbitrary $ \boldsymbol{\hat{\tilde{n}}}$ directions), which reveals the statistical behavior of quantons,

\begin{eqnarray}
\label{eq:20}
&&P(D_0^{'},D_{0}) \equiv P(\alpha \Uparrow_{\tilde{n}},\beta \Uparrow_{\tilde{n}}) \nonumber\\ &&=\frac{1}{4}\lbrack 1-\cos{\vartheta_L}\cos{\vartheta_R} -\sin{\vartheta_L}\sin{\vartheta_R} \cos(2\mu \lambda )\rbrack  \nonumber\\
&&P(D_0^{'},D_{1})\equiv P(\alpha \Uparrow_{\tilde{n}},\beta \Downarrow_{\tilde{n}}) \\ &&=\frac{1}{4}\lbrack 1+\cos{\vartheta_L}\cos{\vartheta_R}+\sin{\vartheta_L}\sin{\vartheta_R} \cos{(2\mu \lambda)}\rbrack \nonumber
\end{eqnarray}

\noindent Then, by introducing observables $\mathcal{O}^{L(R)}$ as in ~\eqref{eq:11}, the expectation values are revealed,

\begin{equation}
\label{eq:21}
    E(\vartheta_L,\vartheta_R) = -\cos{\vartheta_L} \cos{\vartheta_R} -\sin{\vartheta_L} \sin{\vartheta_R} \cos(2\mu \lambda ) 
\end{equation}

\noindent Last, upon substituting ~\eqref{eq:21} into ~\eqref{eq:13}, we derive the correlation function S.

\begin{equation}
\label{eq:22}
\begin{split}
S(\vartheta_L,\vartheta_R,\vartheta_L',\vartheta_R', \lambda) = & \left| -\cos\vartheta_L \cos\vartheta_L' - \sin\vartheta_L \sin\vartheta_L' \cos(2\mu\lambda) \right. \\
& + \left. \cos\vartheta_L \cos\vartheta_R + \sin\vartheta_L \sin\vartheta_R \cos(2\mu\lambda) \right| \\
& + \left| -\cos\vartheta_L' \cos\vartheta_R' - \sin\vartheta_L' \sin\vartheta_R' \cos(2\mu\lambda) \right. \\
& - \left. \cos\vartheta_R \cos\vartheta_R' - \sin\vartheta_R \sin\vartheta_R' \cos(2\mu\lambda) \right|
\end{split}
\end{equation}

\noindent Thus, the correlation function (analogous to Bell angles) for which the retarder phases are maximally violated by the quantum mechanical expectation values is obtained as follows:

\begin{equation}
\label{eq:23}
S(0,\frac{\pi }{4},\frac{3\pi }{4},\frac{\pi}{2},\lambda)=\sqrt{2}+\sqrt{2}\vert \cos (2\mu{\lambda)} \vert 
\end{equation}

\begin{figure}
\centering
\includegraphics[width=0.44\textwidth]{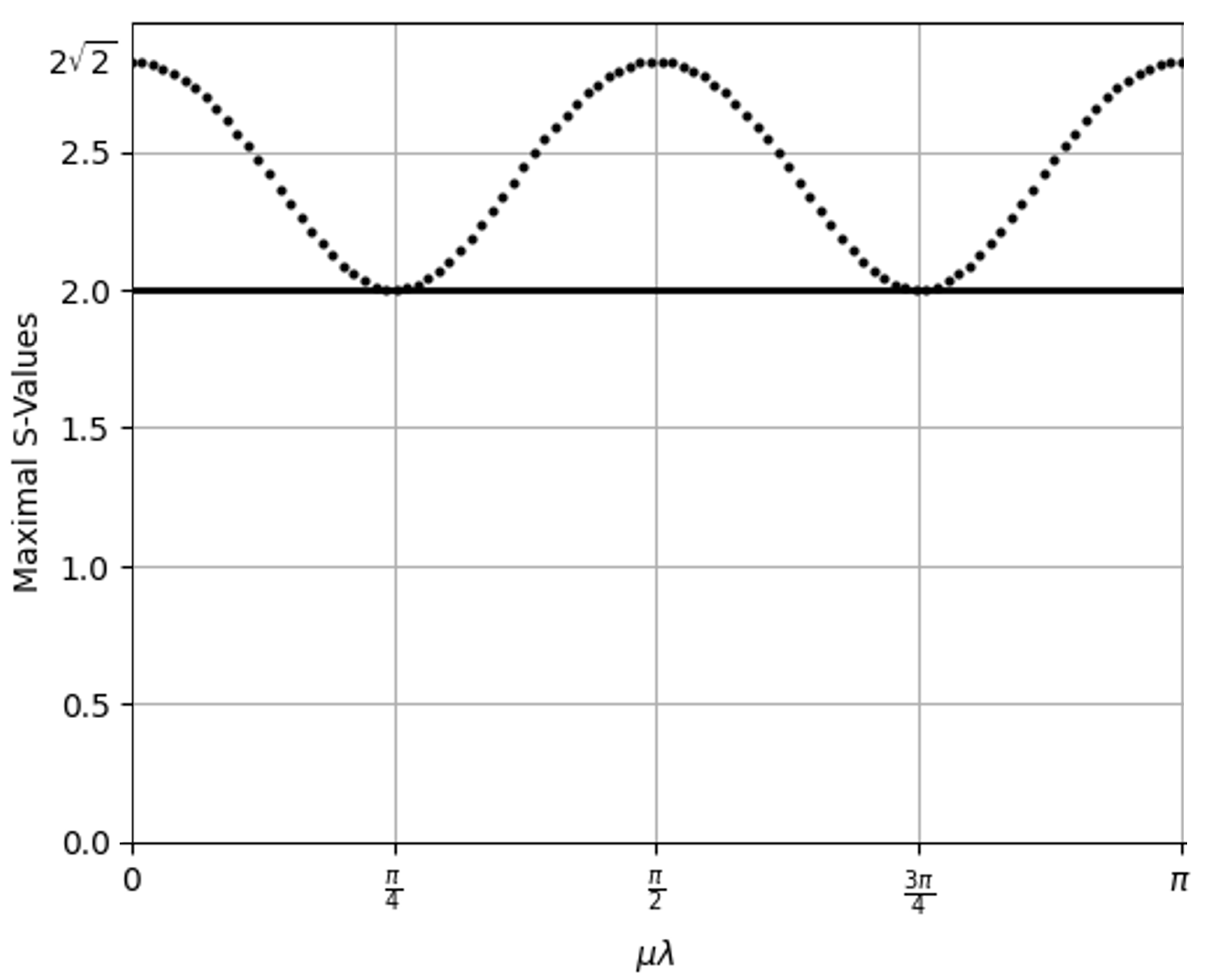}
\caption{\label{fig:wide6}Maximal values of the S function can be explicitly controlled by the AC-HMW phases. Accordingly, without loss of generality, one of the retarder phases can always be eliminated $\vartheta_L=0$. By calculating the extremum conditions of the function \eqref{eq:23} for the other phases of retarders, one can consider the maximal $S$ values for $\mu\lambda$s with the solutions $\vartheta_L'=\pi/ 2$, $\vartheta_R=\pm \arctan[\cos{2\mu\lambda}]$, and $\vartheta_R'= \pi - \vartheta_R$.}
\end{figure}

\noindent This result illustrates how the correlation function depends on the physical quantities in the singular region. Accordingly, for specific angle selections,  $S$ can be controlled by linear charge distributions (See Fig.~\ref{fig:wide6}). In the case of $\lambda=0$, it reduces to $S=2\sqrt{2}$. Besides, when the polarized dipoles are used, there exists entanglement between the spin and path of chargeless particles as its different degrees of freedom. In this case, it is physical noncontextuality rather than the locality that is tested experimentally.

The proposed setup is important for quantum communication and information applications, as the generalizations of BI serve as criteria for entanglement and separability. The nonlocal structure of AC-HMW phases facilitate operations at requisite distances and connections for BI testing, devoid of adiabatic conditions or classical interactions. The robust structure of topological phases offers the possibility of obtaining more precise results due to their resistance to environmental noise. Furthermore, in the setup utilizing spatially correlated particles (with momentum conservation), ~\eqref{eq:23} is valid for all types of quantons (bosons, fermions) that provides flexibility and suitability under diverse laboratory conditions. The use of moving particles as electrically and magnetically neutral dipole carriers is adequate for experimentation ~\cite{Cimmino1989, Gillot2013}. In this context, the setups with neutron interferometers, where the effects of geometric phases on entangled quantum states are examined, stand out as the most inspiring studies ~\cite{Bertlmann2004, Sponar2010, Badurek1988, Hasegawa2003}. Lastly, it should be further emphasized that the results of ~\eqref{eq:23} emerge solely based on the detection probabilities of particles, analogously to spin measurements, and that Bell angles for BI testing can be controlled through phase retarders.

\section{Discussion and conclusions}

Bell-CHSH inequalities (BI) are crucial for testing the fundamental principles of quantum mechanics, particularly in assessing quantum entanglement, one of its most profound aspects within the framework of local realism. Generalizations of BI serve as criteria for entanglement and separability, with experimental implementations conducted in large quantum systems such as photons and atoms. On the other hand, topological AC and HMW phases emerge as excellent tools in developing quantum technologies with their nonlocal features. In this study, our primary motivation is to propose gedanken experimental setups for testing BI utilizing all types of quantons, especially spin 1/2 particles, extending beyond photons. We delve into the effects of spin-dependent AC-HMW phases on entangled quantum states, leveraging their nonlocal structure to gain an experimental advantage. Therefore, we place electromagnetic field sources (linear charge distributions) in the center of BS-P-BS systems, where the motion of particles is not allowed. Through this setup, we observe the topological phase contributions to the wave functions of the moving dipole carrier chargeless particles. Consequently, the correlation function S can be explicitly controlled by the topological phase, thereby influencing the entangled quantum states.

Now, let us have some discussion on our scheme and its possible implementations in physical systems. First, the AC phase is recognized as a special case of the Berry phase that emerges under adiabatic conditions ~\cite{Berry1984, Mignani1991}. In this regard, the research conducted by Bertlmann is significant ~\cite{Bertlmann2004}. In this work, the authors explore the effects of the Berry phase generated by implementing an adiabatically rotating magnetic field into one of the paths of magnetic dipole carrier entangled neutrons moving in opposite directions. They use spin-echo method to eliminate phases that arise due to the system dynamics during the motion of the particles and perform with neutron interferometers. Since their results (See equations (20-22) in ~\cite{Bertlmann2004}) should be reduced to our ~\eqref{eq:23} with Bell angles by restricting the azimuthal angles, it can be reconsidered for the proposed setup in Fig.~\ref{fig:wide5}. In contrast to ~\cite{Bertlmann2004}, our proposed setup offers a significant advantage by utilizing AC-HMW phases, which possess non-local features, without requiring an adiabaticity condition. In this regard, it also clearly reveals the relationship between Berry phase and its special cases AC-HMW phases. Moreover, at this stage, a discussion on noncontextuality needs to be addressed. Noncontextuality requires that observables yield a value that does not depend on the experimental context; results of measurements must be independent of other simultaneous measurements. When the AB effect is considered in two dimensions, the solenoid can be regarded as a dipole and is inherently polarized in the z-direction. Hence, the identity relation between the AB and AC phases holds when AC dipoles are polarized (for $s=\pm1$). In the proposed setup, entanglement occurs between the spin and paths of neutral particles. This suggests experimentally testing physical noncontextuality rather than locality [48-51]. Thus, we use the BI to test noncontextual hidden variable theories.


Second, ~\eqref{eq:23} is also consistent with the findings of [19]. In this study, the authors delve into the AB and AC effects within the framework of fully relativistic quantum mechanics in 2+1 dimensions, demonstrating the ability to control the correlation  the correlation function for BI through the AC phase. To achieve this, they leverage spin measurement probabilities and associated expectation value expressions. Here, we utilize the probabilities of particle detection analogously to spin measurements and use phases of the retarders analogous to Bell angles in the proposed setup. This illustrates the application of Mach-Zender interferometers modified with phase retarders, where the probabilities of detection by the detectors serve as analogs to spin measurements. As a result, this offers an example of conducting entanglement experiments using all kinds of quantons, not limited to photons. Thus, we pave the way for experimental exploration of the effects of phases on entangled quantum states, verification of the spin independence of the AB phase, and examination of the duality and identity relationships between phases, considering the HMW phase as well. Such endeavors are important for quantum communication and information applications.

Third, as a clear example of AC-HMW duality, it reveals the nature of AB-type quantum mechanical effects. It is conceivable that in such an configuration the best known and earliest example of topological phases, the AB phase (which is also the case for the fully dual DAB phase), could be used at the first instance. However, the interaction Hamiltonian describing the dynamics of the AB system and the transformed wavefunction are given by,

\begin{equation}
\label{eq:24}
\begin{matrix}
    \Delta H_{AB} =-e\boldsymbol{\alpha} \cdot\boldsymbol{A} \\
    \psi' =e^{-ie\Phi}\psi_0
\end{matrix}
\end{equation}

\noindent where e is the electron charge, $\boldsymbol{A}$ is the vector potential, and $\Phi$ is the magnetic flux. The phase is independent of the spin orientation of the particles (s), and disappears in the expectation value expressions. The results are reduced to the equations ~\eqref{eq:8} and ~\eqref{eq:9} even in the presence of the solenoid in the case of using AB phase in the proposed setups contrarily done in Silverman's study ~\cite{Silverman1990}. Moreover, although the setup in Fig.~\ref{fig:wide3} works for all types of quantons (bosons, fermions, etc), it is not suitable to investigate the effects of topological phases on entangled quantum states even if we consider the spin-dependent AC-HMW phases. Hence, the probabilities in ~\eqref{eq:7} are related to the joint-spin measurement probabilities in EPR-Bohm-type setups with phase difference $\frac{\pi}{2}$, and since no real closed orbit is formed (single-slit analogy). The configuration in Fig.~\ref{fig:wide4} is well suited to study the effects of nonlocal topological phases (double-slit analogy) and it leads to construction of Fig.~\ref{fig:wide5} ~\cite{Degiorgio1980}. 

Last, although the study can be performed using polarizing BSs, using a different BS each time to perform the spin measurement analogy is not practical for experimental applications. Nevertheless, for the completeness of the study, it can be realized by choosing coefficients r and t for ~\eqref{eq:1} as,

\begin{equation}
\label{eq:25}
BS=\frac{1}{\sqrt{2}}\left(\begin{array}{cc} -\sin{\frac{\vartheta_L}{2}} & -i\sin{\frac{\vartheta_R}{2}} \\
i\cos{\frac{\vartheta_L}{2}} & \cos{\frac{\vartheta_R}{2}} 
\end{array}\right)
\end{equation}

The proposed setup, constructed with spatially correlated particles (via conservation of momentum) analogously to singlet states, does not allow for the examination of mixed states. Nevertheless, conducting such a study is imperative for achieving a thorough understanding of the influence of geometric and topological phases on quantum mechanical states. In this regard, we intend to pursue this line of research in future studies. On the other hand, we believe that modifying the proposed setup to incorporate four entangled quantons will enable the testing of Leggett's inequality, with the aim of exploring the incompatibility between nonlocal realism and quantum mechanics using AC-HMW phases ~\cite{Su2013}. Similarly, it could serve as an experimental setup suggestion for the study with four entangled quantons ~\cite{Pati1998}. 

\section*{Acknowledgements}
The author would like to thank Prof. Ali Ulvi Yilmazer, Prof. Abdullah Verçin, Prof. Anatoli Polkovnikov, Dr. Melik Emirhan Tunalıoğlu, Dr. Tatsuhiko Ikeda and Mary Lynch for their valuable contributions. This work is supported by TUBITAK-BIDEB 2219 Project.

\nocite{*}
\bibliographystyle{elsarticle-num}
\bibliography{refs-2} 

\end{document}